\documentclass[sigconf]{acmart}

\usepackage{graphicx}
\usepackage{booktabs}
\usepackage{amsfonts}
\usepackage{xcolor}
\usepackage{microtype}
\usepackage{enumitem}
\usepackage[most]{tcolorbox}
\usepackage{listings}
\usepackage{placeins}

\lstdefinestyle{compactpython}{
  language=Python,
  basicstyle=\ttfamily\footnotesize,
  keywordstyle=\color{blue!60!black},
  stringstyle=\color{green!45!black},
  commentstyle=\color{gray}\itshape,
  showstringspaces=false,
  breaklines=true,
  breakatwhitespace=true,
  columns=fullflexible,
  keepspaces=true,
  xleftmargin=1em,
  aboveskip=4pt,
  belowskip=4pt
}

\AtBeginDocument{%
  }

\setcopyright{none}
\copyrightyear{2018}
\acmYear{2018}
\acmDOI{XXXXXXX.XXXXXXX}
\acmConference[Conference acronym 'XX]{Make sure to enter the correct
  conference title from your rights confirmation email}{June 03--05,
  2018}{Woodstock, NY}
\acmISBN{978-1-4503-XXXX-X/2018/06}
\renewcommand\footnotetextcopyrightpermission[1]{}

\def\ours{\textsc{SemBaker}}

\begin{document}

\title{From Interpretation to Compilation: A Compilation-Based Execution Engine for Semantic Operator Systems}

\author{Wenkai Dong}
\email{dongw@hawaii.edu}
\affiliation{%
  \institution{University of Hawaii at Manoa}
  \city{Honolulu}
  \state{Hawaii}
  \country{USA}
}

\author{Yifan Wang}
\correspondingauthor
\email{yifanw@hawaii.edu}
\affiliation{%
  \institution{University of Hawaii at Manoa}
  \city{Honolulu}
  \state{Hawaii}
  \country{USA}
}

\begin{abstract}
Semantic operators extend data processing with natural-language predicates. Existing semantic operator systems commonly execute these operators through interpretation-based execution: for every data item, an LLM interprets the operator predicate and directly produces the corresponding result. Although expressive, this design places expensive model invocations inside the data-processing loop, causing latency and monetary cost to scale with input cardinality.

We present SemBaker, a compilation-based execution engine for semantic operator systems. It is not to replace the native execution of a semantic operator system, but to act as an external plugin that allow a bypass of semantic operators to be processed outside the system, which will be faster and cheaper than execution in the system. 
Specifically, we propose ``compilation-based execution'' that uses a lightweight non-LLM program to replace the LLM in runtime interpretation. 
SemBaker lifts this operator-level technique into a complete plan-level engine. Given a semantic operator pipeline, \ours{} uses a cost-model based optimizer to route each operator to either native or compiled execution, and executes mixed plans containing both implementations -- native operators are still executed by the backend system, while compiled ones (programs) are executed locally. This significantly reduces the execution latency and cost, as the non-LLM program is faster and cheaper than LLM call. 
SemBaker currently supports Palimpzest, LOTUS, Nirvana, and DocETL through thin adapters and their existing extension entry-points. 
Our evaluation shows that
\ours{} achieves \(4.8\)--\(6.3\times\) average speedups and
\(5.4\)--\(10.7\times\) average cost reductions with competitive quality.
\end{abstract}

\maketitle

\section{Introduction}
\label{sec:introduction}

Large language models (LLMs) have enabled a new class of data-processing
systems in which users express operations through natural language. For example, a user may filter reviews using the predicate ``the review is clearly positive,'' map each document to its main
topic, or join two records when they describe semantically related entities.
Such operations are commonly exposed as \emph{semantic operators}, including
semantic filter, semantic map, semantic join, and so on. Semantic operator systems~\cite{liu2025palimpzest,patel2025lotus, zhu2025nirvana, shankar2024docetl} make these operators first-class components of declarative data-processing pipelines. By allowing users to specify
semantic intent directly, these systems simplify analytics over text-rich,
semi-structured, and otherwise difficult-to-process data.

The flexibility of semantic operators, however, comes with substantial
execution cost, as they commonly use what we call 
\emph{interpretation-based execution}: the
LLM is invoked at runtime to interpret the semantic operator for each data item (cell or pair of cells).
For example, a semantic filter such as
\texttt{sem\_filter("beginner-friendly machine learning courses")} will be executed by repeatedly sending the predicate together with one row to the LLM
and asking whether that row satisfies the predicate, until all rows are checked. Similarly, a semantic map will send each row to the LLM to obtain a transformed value, while a semantic
join may send each candidate pair of records to the LLM to determine whether the pair satisfies the semantic relationship indicated by the predicate.

This execution model is expressive, but also costly. Each LLM invocation
incurs latency, monetary cost, and possible rate-limit overhead. 
For filters and
maps, the number of LLM calls grows linearly with the number of rows. For joins,
the number of calls grows with the number of candidate pairs, which could be
much larger. As a result, interpretation-based semantic execution can become
prohibitively slow and expensive on large datasets. 

\begin{figure*}[t]
  \centering
  \includegraphics[width=0.9\textwidth]{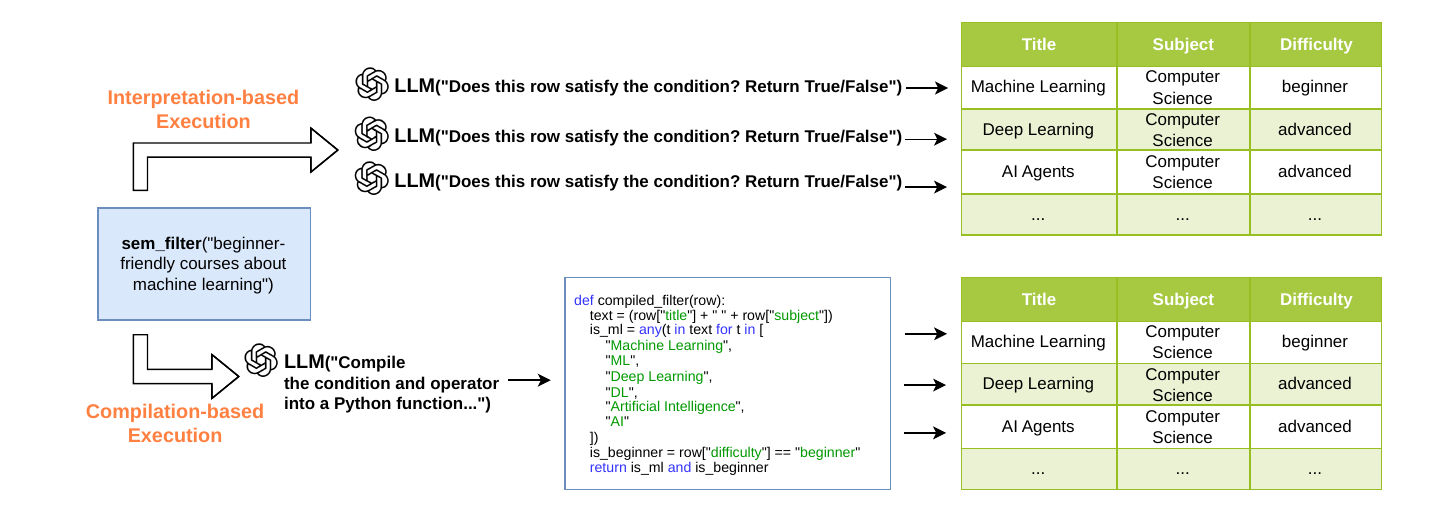}
  \caption{\textbf{Compilation-based execution:} move the LLM from per-row runtime interpretation to
  a one-time, compile-time synthesis of executable operator code. The interpreter
  (top) issues one LLM call per row; the compiler (bottom) issues a single LLM
  call to synthesize a deterministic function (\texttt{compiled\_filter} in the example), which then runs
  locally over every row with no per-row LLM call.}
  \label{fig:overview}
\end{figure*}

Inspired by the idea of compiling SQL into code to speed up query execution in traditional database systems~\cite{hyper}, this paper explores a new execution paradigm for semantic data systems:
\emph{compilation-based execution}. Instead of using an LLM as a runtime
interpreter for every data item, compilation-based execution uses an LLM as a
compiler. 
Given a semantic operator, its natural-language predicate,
schema information, and optional representative samples, the compiler
invokes LLM once to generate deterministic executable code that reproduces the intended behavior of the original semantic operator. The generated program is then executed locally over each data item without additional LLM calls.
Figure~\ref{fig:overview} compares interpretation-based and
compilation-based execution. The former invokes the LLM for each input
item, whereas the latter generates one function and applies it locally
to all inputs. This removes repeated LLM calls and significantly reduces
execution time and cost on large datasets.
Although compilation trades some of the interpreter's flexible reasoning ability for substantially cheaper and more predictable execution, our initial operator-level experiments show that the improvement on efficiency and cost saving is significant while the quality loss is tiny. 
This result motivates a broader systems question: how can compilation-based execution be integrated into existing semantic operator systems without replacing their native execution engines?



We present \ours{}\footnote{\url{https://github.com/iDB-Research-Group/SemBaker}}, a portable compilation-based execution engine for semantic
operator systems. \ours{} is not a new semantic operator system and does not
replace the native execution path of an existing backend. Instead, it acts as
an external plugin that selectively bypasses the backend for operators that can
be executed more efficiently as compiled programs. Operators that are not
selected for compilation continue to execute through the backend's original
implementation.
It introduces
compilation as an additional physical execution option and chooses between
native and compiled execution on an operator-by-operator basis using the optimizer.

Figure~\ref{fig:workflow} summarizes the workflow. Given an operator
pipeline, \ours{} routes each operator to native or compiled execution
and overlaps code generation with pipeline execution. Operators that
are difficult to compile or too small to amortize compilation remain on
the native path. Section~\ref{sec:approach} presents the design.

Thin adapters integrate \ours{} with Palimpzest, LOTUS, Nirvana, and
DocETL~\cite{liu2025palimpzest,patel2025lotus,zhu2025nirvana,
shankar2024docetl} without modifying their source code.


\ours{} currently focuses on three common semantic operators: semantic filter, semantic map, and semantic join. 
These three operators are the major operators in many cases, mirroring selection, projection and join (SPJ) in traditional databases. In the future we will integrate more operators.    

This paper makes the following contributions:
\begin{itemize}[leftmargin=*,nosep]
    \item We introduce a new compilation-based execution paradigm for semantic operators, moving the LLM from per-item runtime interpretation to one-time code synthesis, significantly saving time and cost.
    \item We present \ours{}, an external plugin that enables compilation-based hybrid execution for existing semantic operator systems, effectively enhancing their performance on specific operators. 
    \item We design a cost-model-based optimizer that routes each operator to
    native or compiled execution, and a parallel compilation-execution engine that minimizes the overhead.
    \item We conduct an extensive evaluation, demonstrating
\(4.8\)--\(6.3\times\) average speedups and
\(5.4\)--\(10.7\times\) average cost reductions with competitive quality.
\end{itemize}


\section{SemBaker}
\label{sec:approach}

SemBaker is an external compilation-based execution engine for semantic
operator systems. It does not replace the backend system; instead, it adds
compiled execution as an alternative physical implementation for selected
semantic operators. Operators that remain native are executed by the original
backend, while compiled operators are executed locally by SemBaker without
per-item LLM calls.

\begin{figure*}[t]
  \centering
  \includegraphics[width=0.9\textwidth]{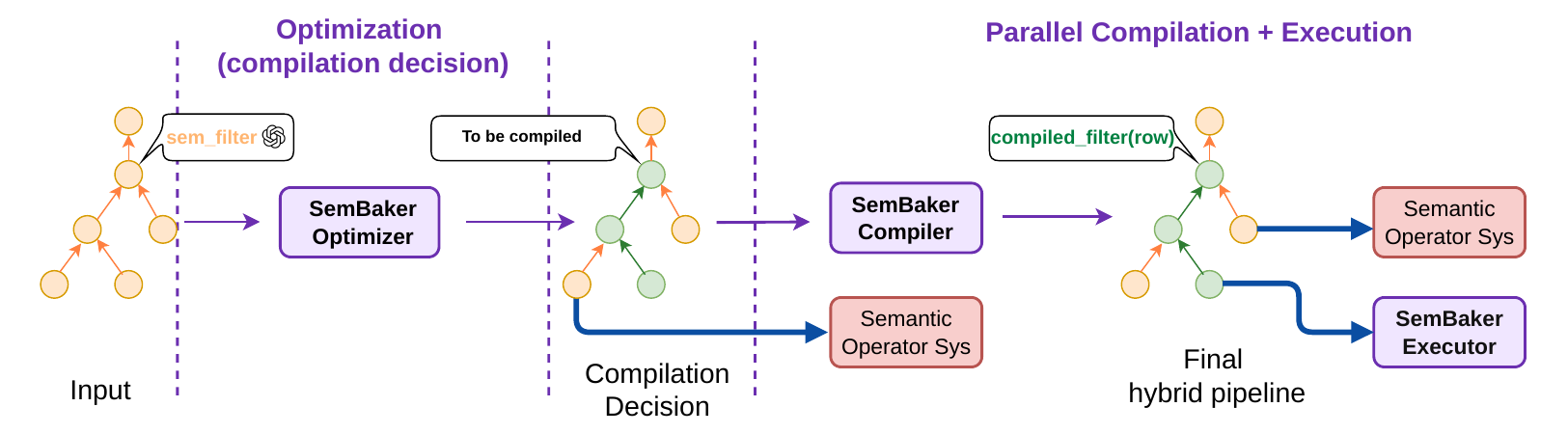}
  \caption{Workflow of \ours{}: The input pipeline consists of all native (LLM-based) operators (orange nodes). It first goes through \ours{} optimizer to decide which operators should be compiled (green nodes), resulting the decision. The decision is then compiled by \ours{} compiler as well as executed by both \ours{} executor and the backend semantic system. The execution starts as soon as the decision is made for all operators, and it only waits if next operator to be executed is not ready. }
  \label{fig:workflow}
\end{figure*}

Figure~\ref{fig:workflow} shows the architecture and workflow of
SemBaker. The system contains three main components: a cost-based \emph{optimizer}, an
LLM-based \emph{compiler}, and a parallel \emph{executor}. Thin backend
adapters connect these components to existing semantic operator systems via their native extension entry points without modifying their source code.
SemBaker also provides helper modules for caching, predicate refinement, and
compiled code validation.
First, the \emph{optimizer} scans a semantic operator plan (\textit{Input} in Figure~\ref{fig:workflow}) and assigns every operator
to either native or compiled execution based on cost estimation, resulting in the \textit{Compilation Decision} in the figure where orange nodes are native and green nodes are to be compiled. This decision is made independently
for each operator, so the input query plan will usually becomes hybrid. 

Then, the \emph{compiler} translates each selected semantic operator into a
deterministic executable program, solidifying the decision to the \textit{final hybrid pipeline} in Figure~\ref{fig:workflow}. The compiler feeds the natural-language predicate, operator type, input
schema, and optional representative records into an LLM, to generate a function that ``mimics'' the operator's behavior/judgment. The generated function is not
allowed to invoke an LLM internally and is reused across all rows or candidate
pairs processed by the operator. SemBaker currently compiles three common
operators:
\[
\begin{aligned}
\texttt{sem\_filter} &\rightarrow
    \texttt{compiled\_filter(row)} \rightarrow \texttt{bool},\\
\texttt{sem\_map} &\rightarrow
    \texttt{compiled\_map(row)} \rightarrow \texttt{value},\\
\texttt{sem\_join} &\rightarrow
    \texttt{compiled\_join(left,right)} \rightarrow \texttt{bool}.
\end{aligned}
\]

Parallelly with the compilation, the \emph{executor} runs the hybrid pipeline. Native operators are sent
to the original backend system. Compiled operators bypass the backend's
LLM-based path and run locally using the generated program. Compilation and
execution are overlapped(parallel), rather than organized as two strictly sequential
phases. The only case where execution waits for compilation is the next operator to be executed is still being compiled. 
Although the compilation and execution are parallel, the execution itself is still in order.  
For example, in Figure~\ref{fig:workflow}, the first native operator (leftmost orange node) can start being executed by the backend semantic system right after decision, at the same time of compilation; while the first compiled operator (bottom green node) has to wait until its code is generated, and the rightmost native operator has to wait until all operators before it are executed.    

\ours{} keeps the optimizer, compiler, executor, and helper modules backend-independent, while uses thin adapters to connect the backends, via their native extension entry points. So \ours{} does not modify any backend's source code, making it portable and extensible. 



\subsection{Compiler}
\label{sec:compiler}

For each operator selected for compilation, SemBaker invokes the LLM once
to generate a deterministic Python function, where each operator has their own compilation prompt.  The prompts are
backend- and dataset-independent: they supply only the operator intent,
input schema, and executable contract.  For example, the default filter
prompt is:

\begin{lstlisting}[
  style=compactpython,
  language={},
  basicstyle=\ttfamily\scriptsize,
  frame=none
]
You are an advanced database query optimizer.
Translate the following semantic predicate into a
highly optimized, deterministic Python function.

Predicate Intent: "{PREDICATE}"

Rules:
1. The function MUST be named `compiled_filter`.
2. It takes one argument `row`, a dictionary with
   the following keys: {INPUT_SCHEMA}.
3. It returns a boolean (True/False).
4. Output ONLY valid Python code, without Markdown
   formatting or explanations.
\end{lstlisting}

Map and join use the same template with operator-specific contracts.
A map returns one value through \texttt{compiled\_map}, whereas a join
returns a Boolean over a record pair through \texttt{compiled\_join}. The join prompt
additionally includes representative unlabeled rows from both inputs so
that the compiler can identify comparable fields.

\textbf{Prompt v1:} 
We also design advanced compilation prompts, Prompt v1. Each retains the complete Default prompt and appends a lightweight natural language processing (NLP) tool calling policy.  
These advanced prompts permit the generated code to use advanced Python NLP and ML packages, like NLTK, NumPy, and scikit-learn. They do not change the compiled function interface, and they prohibit heavy NLP/ML tool calling like LLM calls,
using neural embeddings, network access, model downloads, and building heavyweight neural models.
Prompt V1 is designed for highly semantic queries, allowing more flexibility to the generated code. But it also introduces more compilation and runtime overhead.  

\subsection{Cost-Based Optimizer}
\label{sec:optimizer}

SemBaker makes one routing decision for every supported operators based on its cost model. To estimate the pipeline running cost, the backend adapter first extracts the operator type,
natural-language instruction, visible fields, and an estimate
\(\widehat{n}\) of the amount of work. For filters and maps,
\(\widehat{n}\) is the estimated number of input rows; for joins, it is
the estimated number of candidate pairs. When a materialized input is
available, the adapter obtains its exact cardinality. Otherwise, the cost model
uses such a cardinality estimate: during the plan walk, a sem\_map preserves
cardinality of the input, a sem\_filter propagates an estimated output selectivity, and
a sem\_join multiplies the propagated estimates of its two input branches.
In the current version, we set the selectivity of both
\texttt{sem\_filter} and \texttt{sem\_join} to \(0.5\). Therefore, a
filter with \(n\) input rows is estimated to produce \(0.5n\) rows, and
a join with estimated input cardinalities \(n_l\) and \(n_r\) is
estimated to produce \(0.5n_ln_r\) rows. Note that the amount of work
used to make the join routing decision is still \(n_ln_r\), since all
candidate pairs need to be evaluated. In future work, we will introduce
dynamic selectivity estimation based on the operator predicate and input
data.

Our default optimizer uses a frozen empirical threshold model shared
by all backend adapters. For operator type \(o\) and cost metric \(J\), let \(A_{o,J}\) denote the one-time compilation
overhead and \(r_{o,J}\) the native cost per unit of work (i.e., one row or pair). 
%
Then we have the decision threshold based on amortizing compilation overhead over all work units:
\[
n^{*}_{o,J}
    = \left\lfloor \frac{A_{o,J}}{r_{o,J}} \right\rfloor + 1.
\]
This means when the work unit scale is larger than $n^{*}_{o,J}$, it worth compilation. The default cost metric is processing time. We profile the latency for different operators in our development: the measured profiles use a 25-second compilation overhead and 2.5 seconds per native input row
for filters and maps, giving \(n^{*}=11\). Joins use a 28-second
compilation overhead and 0.15 seconds per native candidate pair,
giving \(n^{*}=187\). The implementation also supports a monetary-cost
metric, whose threshold values are 17, 14, and 17 for filters,
maps, and joins, respectively. These Simple-model profiles are fixed
across backends; backend-specific batching is modeled only by the
optional Advanced cost estimate policy described below.

A threshold alone is insufficient because some predicates cannot be
faithfully implemented by an offline deterministic function. The
default routing rule is therefore
\[
\operatorname{route}(o)=\mathrm{CX}
\quad\Longleftrightarrow\quad
\operatorname{supported}(o)
\land \widehat{n}\ge n^{*}_{o,J}
\land \operatorname{codifiable}(o).
\]
The default codifiability test is a conservative lexical screen for
semantic requirements such as sarcasm, nuanced or subjective
judgment, and multi-hop reasoning -- for highly semantic predicates that are hard to translate to explicit programming logic, they cannot be compiled.  
And unsupported operators are left native.


\subsection{Parallel Compilation and Execution}
\label{sec:pipeline}

SemBaker separates compilation from the backend execution thread and provides a shared scheduler for all adapters. The scheduler accepts a list of compilation tasks and executes them using a bounded thread pool, parallelly with the execution. 
When a pipeline arrives, the optimizer derives compilation tasks, which are
submitted to a persistent background thread pool in plan-execution
order. And execution of the pipeline starts without waiting for full compilation to finish. 
%
%
Native operators invoke the backend's normal
implementation. Compiled operators execute the generated
function once the code becomes available. 
If execution reaches an operator
whose compilation has not finished, it waits only for that operator;
there is no whole-query compilation barrier.
The execution path handles three cases:
(1) If the completed artifact is ready, it is executed immediately.
(2) If it is being compiled, the executor waits until compilation is done, and executes the code. 
(3) Otherwise, if the operator compilation has not yet started, the executor will spawn a compiler thread to compile the operator, store the artifact, and execute it.
Furthermore, the compiled artifacts will be stored in cache, such that they can be reused for duplicate operators to avoid repeated compilation. \ours{} also has retry mechanism to handle compilation failures.    



\subsection{Refine and Validate}
\label{sec:helpers}

Refinement and validation are optional compile-time helpers.
Refinement uses the predicate, schema, and representative unlabeled
records to produce a column-grounded compilation specification. For
example, a join asking for reviews with opposite sentiment can be
rewritten as a deterministic comparison when both inputs contain a
sentiment-label column. The refiner sees no gold labels.

Validation first checks that a generated program is executable and
satisfies its operator contract. It then compares at most three
candidates against LLM-generated pseudo-labels on up to 12 rows for
filters and maps or 20 pairs for joins. SemBaker retains the
highest-scoring executable candidate and stops early when validation
accuracy reaches \(0.8\). Both helpers are disabled by default because
their additional LLM calls do not consistently improve end-to-end
quality.

\section{Experiments}

\subsection{Experimental Setup}

We evaluate \ours{} on all four supported
semantic query engines: Palimpzest (PZ), LOTUS, and Nirvana. 
All experiments use
\texttt{gpt-5-mini-2025-08-07} for both \ours{} and the backends.  Prompt V1, and refinement and validation are enabled only in
their dedicated ablation. They are not integrated in the end-to-end evaluation.

\textbf{Datasets.}
Our main evaluation uses ManyModalQA~\cite{hannan2020manymodalqa},
HybridQA~\cite{chen2020hybridqa}, MultiModalQA
(MMQA)~\cite{talmor2021multimodalqa}, and
SemBench~\cite{lao2025sembench}. The first three are multimodal
question-answering datasets. From SemBench, we use all ten queries in
the movie scenario because they can be implemented consistently across
the three backends and exercise the semantic operators supported by
\ours{}.

Running the complete QA development sets for every backend and method
would be prohibitively expensive. We therefore construct deterministic,
stratified 200-query evaluation sets: 100 text and 100 table questions
from ManyModalQA, 200 hybrid questions from HybridQA, and 50 questions
from each of the four supported non-visual MMQA types. For perspective,
the HybridQA development split alone contains 3,466 questions.
Extrapolating from the observed LOTUS mean latencies of 58.35 seconds
for Native and 45.24 seconds for Ours, a complete Native--Ours pass over
this split would require approximately 100 hours, or 4.2 days, excluding
judge overhead.

We exclude image-dependent questions because the current compiler
targets record-oriented operators over text and tables and does not
consume raw image inputs; such questions would therefore exercise only
the backends' native vision paths. To evaluate compilation when
per-record execution becomes significant, we append 100 deterministic
distractor rows to each table-bearing input in ManyModalQA and MMQA.
The distractors are constructed not to match the reference answers, and
questions whose answers depend on the unconditional table cardinality
are excluded.

We additionally evaluate the DocETL adapter on balanced 1,494-record
subsets of the UCSD Steam Reviews dataset~\cite{wan2018item} for
recommendation mapping and the UCI SMS Spam
Collection~\cite{almeida2011sms} for spam filtering, using 747 records
per class.

Finally, we report the standalone filter, map, and join measurements
from our preliminary study~\cite{dong2026interpretationcompilationcompilationbasedexecution}.

\textbf{Metrics: } For result quality, \textit{Exact Match} (EM) and \textit{F1} score compare the predicted and reference answers directly, computed over successful queries only.
\textit{LLM-based semantic equivalence} (denoted by \textit{LLM} in the evaluation tables) uses an LLM to judge whether an answer is semantically equivalent to the gold answer~\cite{zheng2023judging}.


\subsection{Single-Operator Evaluation}

We first evaluate the performance of compilation-based execution on standalone single-operator. Table~\ref{tab:single-operator} shows one of the results. 
The native implementation serves as the reference,
and \textit{agreement} between compiled and native measures how closely the compiled function reproduces the native operator's
outputs. Since a single operator is not enough to fully answer the queries in the datasets, we did not compare against the groundtruth.  
\textit{Base} is the end-to-end native execution time, while \textit{Comp} and
\textit{Exec} denote one-time compilation and local execution, respectively.
This evaluation shows that compilation preserves most of
the native operators' processing quality and saves time significantly. For example, the compiled functions achieve 0.89--0.96 F1
agreement. Especially, sem\_join attains 0.994 accuracy with up to $31.1\times$ speedup.
More results are in our initial work~\cite{dong2026interpretationcompilationcompilationbasedexecution}. 
\begin{table}[t]
\centering
\small
\setlength{\tabcolsep}{3.5pt}
\caption{Standalone single-operator evaluation. Each operator is evaluated
over 100 predicates. Agreement is measured against the native
implementation. }
\label{tab:single-operator}
\begin{tabular}{l r rrrr rrr}
\toprule
 & & \multicolumn{4}{c}{\textbf{Agreement}}
   & \multicolumn{3}{c}{\textbf{Time per query}} \\
\cmidrule(lr){3-6}\cmidrule(lr){7-9}
\textbf{Operator} & \textbf{\#Q}
& \textbf{Acc.} & \textbf{P} & \textbf{R} & \textbf{F1}
& \textbf{Base (s)} & \textbf{Comp (s)} & \textbf{Exec (ms)} \\
\midrule
Filter & 100 & 0.914 & 0.975 & 0.887 & 0.890 & 130.0 & 28.3 & 1.34 \\
Map    & 100 & 0.912 & 0.911 & 0.919 & 0.911 & 138.7 & 15.5 & 2.43 \\
Join   & 100 & 0.994 & 0.970 & 0.980 & 0.961 & 703.4 & 22.6 & 4.23 \\
\bottomrule
\end{tabular}
\end{table}

\begin{table*}[!t]
  \caption{End-to-end results. QA workloads contain 200 queries,
SemBench contains all 10 movie queries, and each DocETL task contains
1,494 records. Time and cost are per query for QA and SemBench and per
job for DocETL. Judge overhead is excluded; refinement and validation
are disabled.}
  \label{tab:pipeline-main}

  \centering
  \footnotesize
  \renewcommand{\arraystretch}{1.08}
  \setlength{\tabcolsep}{2.5pt}

  \begin{tabular*}{0.96\textwidth}{
    @{\extracolsep{\fill}}
    llrrrrrrrrrr
    @{}
  }
    \toprule
    & &
    \multicolumn{2}{c}{\textbf{EM/Acc.}} &
    \multicolumn{2}{c}{\textbf{F1}} &
    \multicolumn{2}{c}{\textbf{LLM}} &
    \multicolumn{2}{c}{\textbf{Time (s)}} &
    \multicolumn{2}{c}{\textbf{Cost (\$)}} \\
    \cmidrule(lr){3-4}
    \cmidrule(lr){5-6}
    \cmidrule(lr){7-8}
    \cmidrule(lr){9-10}
    \cmidrule(lr){11-12}

    \textbf{Dataset} &
    \textbf{Engine} &
    \textbf{Native} & \textbf{Ours} &
    \textbf{Native} & \textbf{Ours} &
    \textbf{Native} & \textbf{Ours} &
    \textbf{Native} & \textbf{Ours} &
    \textbf{Native} & \textbf{Ours} \\
    \midrule

    ManyModalQA
      & PZ
      & 0.633 & 0.568
      & 0.773 & 0.698
      & 0.920 & 0.844
      & 159.05 & 13.72
      & 0.0184 & 0.0031 \\
      & LOTUS
      & 0.337 & 0.403
      & 0.465 & 0.534
      & 0.942 & 0.942
      & 16.88 & 15.63
      & 0.0286 & 0.0044 \\
      & Nirvana
      & 0.608 & 0.610
      & 0.729 & 0.749
      & 0.901 & 0.900
      & 57.56 & 19.04
      & 0.0274 & 0.0036 \\
    \midrule

    HybridQA
      & PZ
      & 0.455 & 0.467
      & 0.544 & 0.576
      & 0.600 & 0.658
      & 173.28 & 46.33
      & 0.1875 & 0.0147 \\
      & LOTUS
      & 0.391 & 0.435
      & 0.525 & 0.584
      & 0.706 & 0.745
      & 58.35 & 45.24
      & 0.1657 & 0.0202 \\
      & Nirvana
      & 0.399 & 0.430
      & 0.483 & 0.532
      & 0.561 & 0.615
      & 661.48 & 50.68
      & 0.2789 & 0.0180 \\
    \midrule

    MMQA
      & PZ
      & 0.490 & 0.415
      & 0.555 & 0.467
      & 0.630 & 0.520
      & 420.01 & 38.20
      & 0.0319 & 0.0029 \\
      & LOTUS
      & 0.200 & 0.155
      & 0.263 & 0.214
      & 0.640 & 0.580
      & 23.97 & 36.12
      & 0.0820 & 0.0120 \\
      & Nirvana
      & 0.405 & 0.430
      & 0.456 & 0.490
      & 0.500 & 0.540
      & 84.57 & 31.33
      & 0.0610 & 0.0102 \\
    \midrule

    SemBench
      & PZ
      & 0.896 & 0.975
      & 0.917 & 1.000
      & -- & --
      & 68.13 & 18.37
      & 0.0184 & 0.0023 \\
      & LOTUS
      & 0.928 & 1.000
      & 0.934 & 1.000
      & -- & --
      & 8.20 & 20.42
      & 0.0233 & 0.0042 \\
      & Nirvana
      & 0.882 & 0.800
      & 0.793 & 0.586
      & -- & --
      & 43.59 & 11.78
      & 0.0326 & 0.0029 \\
    \midrule

    Steam
      & DocETL
      & 0.870 & 0.686
      & 0.873 & 0.679
      & -- & --
      & 67.98 & 56.40
      & 1.0529 & 0.0083 \\

    SMS
      & DocETL
      & 0.977 & 0.950
      & 0.977 & 0.948
      & -- & --
      & 117.53 & 78.23
      & 2.1733 & 0.0119 \\

    \bottomrule
  \end{tabular*}
\end{table*}

\subsection{End-to-end Results}


Table~\ref{tab:pipeline-main} compares the native engine with \ours{} on the end-to-end performance.  Native and Ours are shown in separate
subcolumns for every metric.
The central result is that our method achieves significant acceleration and token cost saving with a competitive end-to-end processing quality against the native execution.  For instance, on ManyModalQA and HybridQA, it reduces average latency by 79.3\% and 84.1\%, and cost by
85.1\% and 91.6\%, respectively, while keeping quality close to
or even above the native baseline.

\begin{table*}[!t]
\caption{Ablations using a shared baseline: Default prompt, cost-based
routing, and disabled refinement and validation. Each quality cell
reports EM/Accuracy, F1, and semantic-judge score; each efficiency cell
reports time and cost.}
\label{tab:ablations}
\centering
\scriptsize
\setlength{\tabcolsep}{1.8pt}
\renewcommand{\arraystretch}{0.92}

\begin{tabular*}{\textwidth}{
  @{\extracolsep{\fill}}
  ll
  cc
  cc
  cc
  cc
  @{}
}
\toprule
& &
\multicolumn{2}{c}{\textbf{Shared Baseline}} &
\multicolumn{2}{c}{\textbf{Prompt v1}} &
\multicolumn{2}{c}{\textbf{All-CX}} &
\multicolumn{2}{c}{\textbf{Ref.+Val.}} \\
\cmidrule(lr){3-4}
\cmidrule(lr){5-6}
\cmidrule(lr){7-8}
\cmidrule(lr){9-10}

\textbf{Dataset}
& \textbf{Engine}
& \shortstack{\textbf{Quality}\\\textbf{EM/F1/J}}
& \shortstack{\textbf{Efficiency}\\\textbf{Time/Cost}}
& \shortstack{\textbf{Quality}\\\textbf{EM/F1/J}}
& \shortstack{\textbf{Efficiency}\\\textbf{Time/Cost}}
& \shortstack{\textbf{Quality}\\\textbf{EM/F1/J}}
& \shortstack{\textbf{Efficiency}\\\textbf{Time/Cost}}
& \shortstack{\textbf{Quality}\\\textbf{EM/F1/J}}
& \shortstack{\textbf{Efficiency}\\\textbf{Time/Cost}} \\
\midrule

ManyModalQA
& PZ
& 0.568/0.698/0.844 & 13.72/0.0031
& 0.600/0.712/0.855 & 14.42/0.0031
& 0.520/0.635/0.760 & 26.40/0.0057
& 0.568/0.692/0.824 & 17.07/0.0078 \\

& LOTUS
& 0.403/0.534/0.942 & 15.63/0.0044
& 0.398/0.524/0.923 & 21.05/0.0053
& 0.370/0.486/0.920 & 29.40/0.0063
& 0.349/0.481/0.954 & 18.93/0.0087 \\

& Nirvana
& 0.610/0.749/0.900 & 19.04/0.0036
& 0.615/0.758/0.920 & 16.93/0.0049
& 0.620/0.749/0.880 & 25.50/0.0052
& 0.615/0.753/0.890 & 22.86/0.0100 \\

\addlinespace[0.35ex]
HybridQA
& PZ
& 0.467/0.576/0.658 & 46.33/0.0147
& 0.460/0.586/0.662 & 46.06/0.0161
& 0.490/0.620/0.660 & 36.00/0.0146
& 0.390/0.474/0.530 & 89.19/0.0596 \\

& LOTUS
& 0.435/0.584/0.745 & 45.24/0.0202
& 0.397/0.555/0.734 & 49.95/0.0201
& 0.420/0.568/0.730 & 43.20/0.0145
& 0.383/0.515/0.694 & 73.86/0.0591 \\

& Nirvana
& 0.430/0.532/0.615 & 50.68/0.0180
& 0.385/0.485/0.545 & 47.70/0.0195
& 0.480/0.591/0.670 & 35.30/0.0097
& 0.390/0.486/0.575 & 77.16/0.0449 \\

\addlinespace[0.35ex]
MMQA
& PZ
& 0.425/0.475/0.530 & 50.33/0.0030
& -- & --
& 0.290/0.326/0.355 & 34.26/0.0089
& 0.446/0.500/0.549 & 58.94/0.0170 \\

& LOTUS
& 0.177/0.242/0.591 & 48.76/0.0123
& -- & --
& 0.151/0.202/0.508 & 63.62/0.0117
& 0.132/0.186/0.574 & 70.48/0.0260 \\

& Nirvana
& 0.415/0.466/0.525 & 47.46/0.0101
& -- & --
& 0.415/0.444/0.485 & 48.08/0.0094
& 0.425/0.476/0.530 & 59.59/0.0227 \\

\addlinespace[0.35ex]
SemBench
& PZ
& 0.975/1.000/-- & 18.37/0.0023
& 0.563/0.500/-- & 18.24/0.0016
& -- & --
& 0.738/0.500/-- & 32.11/0.0105 \\

& LOTUS
& 1.000/1.000/-- & 20.42/0.0042
& 0.990/0.990/-- & 25.75/0.0043
& -- & --
& 1.000/1.000/-- & 37.07/0.0153 \\

& Nirvana
& 0.800/0.586/-- & 11.78/0.0029
& 0.825/0.586/-- & 16.38/0.0030
& -- & --
& 1.000/0.836/-- & 19.55/0.0147 \\

\addlinespace[0.35ex]
Steam
& DocETL
& 0.711/0.698/-- & 43.96/0.0078
& -- & --
& -- & --
& 0.641/0.646/-- & 84.54/0.0125 \\

SMS
& DocETL
& 0.950/0.948/-- & 78.23/0.0119
& 0.927/0.928/-- & 57.41/0.0106
& -- & --
& 0.901/0.897/-- & 46.49/0.0183 \\

\bottomrule
\end{tabular*}
\end{table*}

\subsection{Ablation Studies}
\label{sec:ablations}

We ablate three design choices in SemBaker: compiler guidance, selective
routing, and refinement with validation.  All variants share the same
baseline, which uses the default compiler profile, cost-based routing,
and turned off refinement and validation. \textit{Prompt v1} method replaces the default compiler prompt with the v1 prompt; \textit{All-CX} method forces compiling every supported
semantic operator; and Ref.+Val. enables refine and validate
modules. Table~\ref{tab:ablations} reports the results.  

\textbf{Compiler guidance needs joint design: }
Prompt v1 helps when the operator semantics align closely with its lightweight
NLP primitives. This is visible on ManyModalQA + PZ and Nirvana, as well as SemBench + Nirvana, where Prompt v1 achieves quality gains than default prompt. However, it is unstable for
operators that require contextual judgment or precise schema reasoning, like on PZ + 
SemBench, F1 decreases from 1.00 to 0.50.  Its efficiency effect is also
conditional.  V1 accelerates SMS by 26.6\%, but increases LOTUS latency on
ManyModalQA.  This evaluation shows that additional primitives change the
types of semantics that the compiler can represent well, indicating that further improving \ours{} needs a systematic design jointly considering all the three major components.    

\textbf{Selective routing is necessary:}
 All-CX performs well when a workload is dominated by semantic operations that
can be captured faithfully by deterministic code.  HybridQA illustrates this
where the queries are relatively less semantic than other datasets: forcing compilation improves F1 on PZ and Nirvana while also reducing
their latency.  In contrast, forced compilation substantially harms workloads
with more complex semantics: On PZ, All-CX reduces F1 by 14.9\% on MMQA and also produces a clear quality and latency regression on
ManyModalQA. These results prove the necessity of \ours{} optimizer and selective routing that let the execution benefit from both compilation time/cost-saving and LLM's flexible reasoning. 

\paragraph{Refinement and validation are a trade-off: }
Ref.+Val. significantly improve the quality like on SemBench + Nirvana, but they may also decrease the quality in some cases, meaning that they are unstable in different cases.  
This result highlights a limitation of adding
LLM-based quality-control stages indiscriminately: refinement may alter the
original intent, while validation based on a small sample may not reliably
predict full-workload behavior. This further strengths that the further improvement of \ours{} needs a joint design.  

\textbf{Overall conclusion:}
Overall, the current architecture of \ours{} provides the best trade-off
among the evaluated configurations under the current scenarios. Further
improvement requires a joint design of the whole system rather than an
upgrade to a single component.

\section{Conclusion}
\label{sec:conclusion}

We presented \ours{}, an external compilation-based engine that adds
deterministic local execution to existing semantic operator systems.
Through operator routing and parallel compilation and execution, \ours{} achieves significant acceleration and cost saving without replacing or
modifying the backend systems. In evaluation, 
\ours{} achieves \(4.8\)--\(6.3\times\) average speedups and
\(5.4\)--\(10.7\times\) average cost reductions, with competitive quality to the backend systems.

\FloatBarrier
\bibliographystyle{ACM-Reference-Format}
\bibliography{main}

\end{document}